\documentclass[aps,twocolumn,superscriptaddress,amsfont,amssymb,amsmath,nofootinbib,showpacs,balancelastpage,english]{revtex4-1} %\documentclass[aps,prd,preprint,preprintnumbers,notitlepage,amssymb,preprintnumbers,notitlepage,nofootinbib,10pt,english]{revtex4}
\usepackage{amsmath}
\usepackage{amssymb}
\usepackage{babel}
\usepackage{graphicx}
\usepackage{color}
\usepackage{txfonts}
\usepackage{natbib}
\usepackage{mathrsfs}
\usepackage{longtable}



\begin{document}

\title{Measuring Speed of Gravitational Waves by Observations of Photons and Neutrinos from Compact Binary Mergers and Supernovae}

\author{Atsushi Nishizawa}
\email{anishi@caltech.edu}
\affiliation{Theoretical Astrophysics 350-17, California Institute of Technology, Pasadena, California 91125, USA}
\author{Takashi Nakamura}
\affiliation{Department of Physics, Kyoto University, Kyoto 606-8502, Japan}

\begin{abstract}%%%%%%%%%%%%%%%%%%%%%%%%%%%%%%%%%%%%%%%%%
Detection of gravitational waves (GW) provides us an opportunity to test general relativity in strong and dynamical regimes of gravity. One of the tests is checking whether GW propagates with the speed of light or not. This test is crucial because the velocity of GW has not ever been directly measured. Propagation speed of a GW can deviate from the speed of light due to the modification of gravity, graviton mass, and the nontrivial spacetime structure such as extra dimensions and quantum gravity effects. Here we report a simple method to measure the propagation speed of a GW by directly comparing arrival times between gravitational waves, and neutrinos from supernovae or photons from short gamma-ray bursts. As a result, we found that the future multimessenger observations of a GW, neutrinos, and photons can test the GW propagation speed with the precision of $\sim 10^{-16}$, improving the previous suggestions by 8-10 orders of magnitude. We also propose a novel method that distinguishes the true signal due to the deviation of GW propagation speed from the speed of light and the intrinsic time delay of the emission at a source by looking at the redshift dependence. 
\end{abstract}

\date{\today}

\maketitle

%%%%%%%%%%%%%%%%%%%%%%%%%%%%%%%%%%%%%%%%%%%%%

\section{Introduction}
Direct detection experiments of a gravitational wave (GW) have been well developed in a past decade. In a few years, the next-generation kilometer-scale laser-interferometric GW detectors such as aLIGO \cite{Harry:2010zz}, aVIRGO \cite{Accadia:2011zzc}, and KAGRA \cite{Somiya:2011np} will start observations and are expected to achieve the first detection of a GW. The GW observations enable us to test gravity theory in strong and dynamical regimes of gravity (for reviews, see \cite{Will:2014kxa,Yunes:2013dva,Gair:2012nm}). There have been the suggestions of model-independent method to test gravity by searching for anomalous deviation in the GW phase predicted in general relativity \cite{Mishra:2010tp,DelPozzo:2011pg,Li:2011cg,Yunes:2009ke,Yunes:2010qb,Cornish:2011ys} and with GW polarizations \cite{Nishizawa:2009bf,Nishizawa:2009jh,Chatziioannou:2012rf,Hayama:2012au}. The other test is measuring the propagation speed of a GW. In general relativity, a GW propagates with the speed of light, while in the alternative theories of gravity the propagation speed could deviate from the speed of light due to the modification of gravity (see for general formulation \cite{Saltas:2014dha}, and for more specific cases, nonzero graviton mass \cite{Gumrukcuoglu:2011zh,DeFelice:2013nba} and extra dimensions \cite{Sefiedgar:2010we}). Also the modification of spacetime structure due to quantum gravity effects may affect the propagation of a GW \cite{AmelinoCamelia:1997gz,AmelinoCamelia:2008qg}. 

Recently BICEP 2 team announced the first detection of GWs in the B-mode polarization data \cite{Ade:2014xna}. However, there is the possibility that the signal is the contamination by polarized dust emission \cite{Mortonson:2014bja,Flauger:2014qra} and the detection is still controversial. If the detection is true, the speed of a GW is measured \cite{Raveri:2014eea,Amendola:2014wma}, though the constraint is weak with the present data. On the other hand, the propagation speed of a GW has been constrained from the observations of ultra-high energy cosmic rays. If GW velocity is subluminal, then cosmic rays lose their energy via gravitational Cherenkov radiation and cannot reach the Earth. The fact that ultra-high energy cosmic rays are observed on the Earth limit the GW propagation speed to be $c-\upsilon_g < 2 \times 10^{-15} c$, assuming the cosmic rays have galactic origin \cite{Moore:2001bv}. However, superluminal GW is not constrained at all with this observation. In either case, since the velocity of a GW has not ever been directly measured, the accurate measurement of the GW propagation speed in direct detection experiment is crucial in testing the gravity theories. 

So far there are some proposals to directly measure the propagation speed of a GW. One is the method with the R\o mer time delay \cite{Finn:2013toa}. A GW signal from a periodic GW source is modulated in phase due to the Earth revolution. This method relies only on GW observation and is a direct measurement of the GW propagation speed. However, the sensitivity is limited by the relatively small baseline of the propagation, $\sim 1{\rm{AU}}$. Another method is comparing the phases of a GW and its electromagnetic counterpart from a periodic source \cite{Larson:1999kg,Cutler:2002ef}. The authors have considered a white dwarf binary as a periodic source. However, we do not know the intrinsic phase lag between GWs and electromagnetic waves at the times of emissions. For the cancellation of the unknown phase, two signals at different times (e.g. a half year) have to be differentiated, which again limits the signal gain to $\sim 1\,{\rm{AU}}$ baseline. 

In this paper, we report a simple method measuring the propagation speed of a GW by directly comparing the arrival times between gravitational waves, and neutrinos or photons from short gamma-ray burst (SGRB) and supernovae (SN). Here we assume that SGRB is associated with a NS-NS  or NS-BH binary merger (see for example the recent review \cite{Berger:2013jza}), where NS and BH mean neutron star and black hole, respectively. In comparing the arrival times, the uncertainty of the intrinsic time delay of GWs, neutrinos, and photons prevent us from interpreting the difference in arrival times as that in propagation times. However, for SGRB and SN, their model buildings and the numerical simulations have been well developed in these days and start to allow us to discuss the intrinsic time delays. We will show that using the input from the numerical simulations, we can tightly constrain the propagation speed of a GW with significantly improved sensitivity than previous studies. 

\begin{figure}[t]
\begin{center}
\includegraphics[width=8cm]{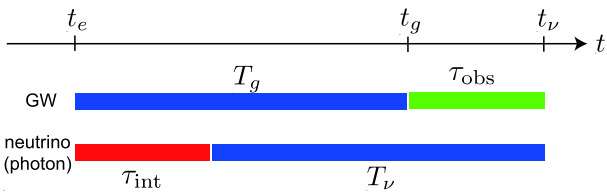}
\caption{GW and neutrino (photon) propagation times. GW is emitted at the time $t=t_e$ and detected on the Earth at $t=t_g$. For instance, we refer the merger time of a NS binary or the core bounce time of a core-collapsed SN to the emission time of a GW, while a neutrino (photon) is emitted at $t=t_e+\tau_{\rm{int}}$ and detected at $t=t_{\nu}$. The observable is the difference of the arrival times between the GW and neutrinos(photons), $\tau_{\rm{obs}}$}
\label{fig1}
\end{center}
\end{figure}

%%%%%%%%%%%%%%%%%%%%%%%%%%%%%%%%%%%
\section{Method}
In general, a massive particle has dispersion relation, $E^2=m^2+p^2$. For the relativistic regime $m \ll E$, the propagation speed (group velocity) is 
\begin{equation}
\upsilon \equiv \frac{\partial E}{\partial p} \approx c \left(1-\frac{m^2 c^4}{2E^2}\right) \;. 
\label{eq6}
\end{equation}
Then the propagation time from a source at the distance $L$ is given by
\begin{equation}
T =\frac{L}{\upsilon} \approx T_0 \left(1+ \frac{m^2c^4}{2E^2} \right) \;,
\label{eq5}
\end{equation}
where $T_0 \equiv L/c$. Note that for cosmological sources we must use 
the exact formula of the distance that takes into account cosmic expansion. However, as long as we consider sources at low redshift ($z<0.1$), Eq.~(\ref{eq5}) causes the error less than $5\%$. When we discuss the sources at cosmological distance in Sec.~\ref{sec4}, we use the exact formula.

Let us start from the comparison of the propagation speeds of a GW and neutrinos. To do so, we write the lightest mass of neutrinos among three mass eigenstates of neutrinos as $m_{\nu}$ and define the fastest propagation speed of neutrinos as
\begin{equation}
\upsilon_{\nu} = c\left( 1-\frac{m_{\nu}^2c^4}{2E_{\nu}^2}\right) \;, 
\end{equation}
where $E_{\nu}$ is the energy of the neutrino. As in Fig.\ref{fig1}, a GW is emitted at the time $t=t_e$ and detected on the Earth at $t=t_g$ (for instance, we refer the merger time of a NS binary or the core bounce time of a core-collapsed supernova to the emission time of a GW), while a neutrino (photon) is emitted at $t=t_e+\tau_{\rm{int}}$ and detected at $t=t_{\nu}$. The observable is the difference of the arrival times between the GW and neutrinos(photons), $\tau_{\rm{obs}}$ in Fig.~\ref{fig1}, which is positive (negative) for an early (late) arrival of a GW. Usually the emission times of a GW and neutrinos at a source do not coincide so that $\tau_{\rm{obs}}$ contains the intrinsic time delay at a source $\tau_{\rm{int}}$. In some theories of modified gravity, a GW and other particles couple to different effective metrics and then the Shapiro delays during propagation are different \cite{Desai:2008vj,Kahya:2010dk}. However, we exclude the case in the following discussion. Denoting the propagation times of GW and neutrinos, $T_g \equiv L/\upsilon_g$ and $T_{\nu} \equiv L/\upsilon_{\nu}$, respectively, and defining $\Delta T \equiv T_{\nu} - T_g$, we can express the difference of arrival times as  
\begin{equation}
\tau_{\rm{obs}} =\Delta T + \tau_{\rm{int}} \;.
\label{eq12}
\end{equation}
The first term of Eq.~(\ref{eq12}) contains the time lags due to the possible deviation of the GW propagation speed from the speed of light and the contribution of non-zero neutrino mass. Under the assumption that a GW propagates with the speed of light, the method to measure neutrino mass has been discussed in \cite{Arnaud:2001gt}, though the mass detection would be difficult due to the smallness of the mass, as we discuss later. The second term comes from the intrinsically delayed emission time of neutrinos at a source. 

In order that the finite time lag due to the GW propagation speed different from the speed of light and neutrino mass is detectable, $\Delta T$ has to exceed uncertainties in the intrinsic time lag of the emissions and satisfy one of the following two conditions:
\begin{align}
&\Delta T + \tau_{\rm{int,max}} < \tau_{\rm{int,min}} \quad \quad {\rm{for}}\;\;\Delta T <0 \;, \\
&\tau_{\rm{int,max}} <  \Delta T + \tau_{\rm{int,min}} \quad \quad {\rm{for}}\;\;\Delta T >0\;,
\end{align}
equivalently, 
\begin{equation}
\Delta \tau_{\rm{int}} <  |\Delta T| \;,
\label{eq15a}
\end{equation}
with $\Delta \tau_{\rm{int}} \equiv \tau_{\rm{int,max}}-\tau_{\rm{int,min}}$.

It is convenient to define the deviation from the speed of light as $\delta_g \equiv (c-\upsilon_g)/c$ and $\delta_{\nu} \equiv (c-\upsilon_{\nu})/c$. Expressing $\Delta T$ in terms of $\delta_g$ and $\delta_{\nu}$ and keeping the leading order in $\Delta T/T_{\nu}$ give the relation
\begin{equation}
\frac{\Delta T}{T_0} \approx \delta_{\nu}-\delta_g \;. \label{eq3}
\end{equation}
Substituting this into Eq.~(\ref{eq15a}), we obtain
\begin{equation}
\Delta \tau_{\rm{int}} < T_0 |\delta_{\nu} - \delta_g| \;,
\label{eq15}
\end{equation}
with
\begin{equation}
\delta_{\nu} = \frac{m_{\nu}^2c^4}{2E_{\nu}^2} \;. \label{eq1}
\end{equation}
Thus depending on the uncertainty in the intrinsic time delay and neutrino mass, we can detect the deviation of the GW propagation speed from the speed of light, $\delta_g$. For the comparison of the propagation speeds between a GW and photons, the detectable range of $\delta_g$ is obtained by merely setting $\delta_{\nu}=0$ in Eq.~(\ref{eq15}) since a photon is massless. 

In Eq.~(\ref{eq15}), we have not taken into account the timing errors of a GW, neutrinos, and photons when detected on the Earth. However, the detection timing error can be neglected because the intrinsic uncertainty of the emission time is much larger, as we discussed later (e.g. $\sim 10^{-2}\,{\rm{sec}}$ for SN neutrinos and $\sim 10\,{\rm{sec}}$ for SGRB photons). On the other hand, the phase error of a GW significantly depends on the signal-to-noise ratio (SNR) and is given by $\Delta \phi_{\rm{gw}} \sim {\cal{O}}({\rm{SNR}})^{-1}$ \cite{Cutler:1994ys}. In the case of GW detection from SN and SGRB at $\sim 100\,{\rm{Hz}}$ with a ground-based detector such as aLIGO, SNR is $\sim10$ for SN at the distance of $100\,{\rm{kpc}}$ and SGRB at $200\,{\rm{Mpc}}$. For these sources, the detection timing error of a GW is at most $\sim 10^{-3}\,{\rm{sec}}$. Therefore, the detection timing errors can be neglected when we consider the constraint on $\delta_g$ from the typical sources of a GW with electromagnetic or neutrino counterparts.

%%%%%%%%%%%%%%%%%%%%%%%%%%%%%%%%%%%
\section{Constraint on GW propagation speed}
\label{sec3}

First let us estimate what parameter ranges we can detect in the deviation of the GW propagation speed from the speed of light and the lightest neutrino mass with the GW-SN multimessenger observation.

Most numerical simulations of SN with rotating progenitors predict that neutrinos are emitted within $10^{-2}\,{\rm{sec}}$ after the core bounce and that GWs are mainly radiated sharply at the time of the core bounce \cite{Ott:2012kr,Kuroda:2013rga}. On the other hand, this is not the case for non-rotating collapses. The GW emission follows the neutrino emission because the development of turbulence is delayed \cite{Marek:2008qi,Kuroda:2013rga}. However, the GW waveform of the non-rotating core-collapse could easily be distinguished by the GW observation since the GW waveform does not accompany characteristic spikes at the time of the core bounce. From this reason, we focus on the SN with rotating progenitors and consider the intrinsic time delay of neutrino emission to be at most $10\,{\rm{msec}}$. 

\begin{figure}[t]
\begin{center}
\includegraphics[width=8.5cm]{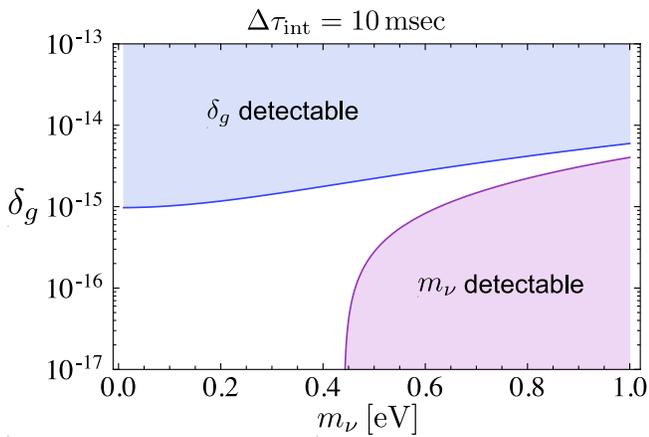}
\caption{Detectable parameter ranges of the deviation of GW propagation speed (when $\delta_g>0$) and lightest neutrino mass from the multimessenger observation of a GW and SN neutrinos. The neutrino energy is $10\,{\rm{MeV}}$ and the distance to the source is $100\,{\rm{kpc}}$.}
\label{fig2a}
\end{center}
\end{figure}

\begin{figure}[t]
\begin{center}
\includegraphics[width=8.5cm]{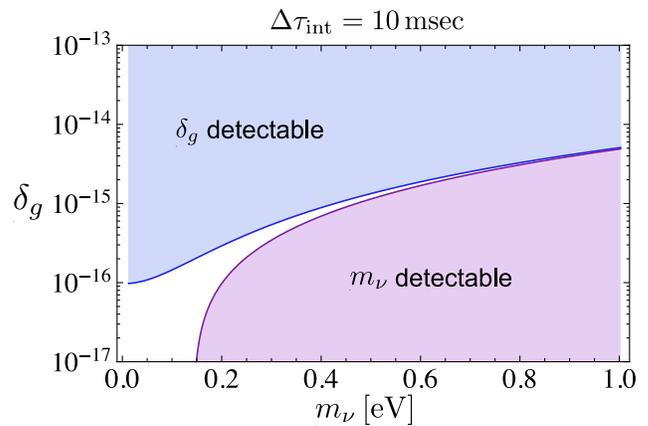}
\caption{Same as Fig.~\ref{fig2a} but the source distance is $1\,{\rm{Mpc}}$.}
\label{fig2b}
\end{center}
\end{figure}

Using Eqs.~(\ref{eq1}) and (\ref{eq15}), we find the detectable parameter ranges of the deviation of GW propagation speed from the speed of light and the lightest neutrino mass from the multimessenger observations of a GW and SN neutrinos. We first consider the subluminal case with $\delta_g>0$, for which $\delta_g$ and $\delta_{\nu}$ cancel at certain parameters since both quantities are positive. Figure~\ref{fig2a} is shown for a SN source located at the distance of $100\,{\rm{kpc}}$, which is the detection range of a gravitational wave detector such as aLIGO and current neutrino detectors. In the figure, we assume that the neutrino energy is $10\,{\rm{MeV}}$ and that the intrinsic time delay is $\Delta \tau_{\rm{int}}=10\,{\rm{msec}}$. If the lightest neutrino mass is $m_{\nu}<0.3\,{\rm{eV}}$, the neutrino mass effect can be neglected and the detectable range of $\delta_g$ hardly changes. Since the constraint obtained by the recent cosmological observations is $\sum m_{\nu} \leq 0.34\,{\rm{eV}}$ (95\% CL) in the flat $\Lambda$CDM model \cite{Zhao:2012xw}, we can ignore the neutrino mass uncertainty in constraining $\delta_g$. If next-generation detectors of GWs (Einstein Telescope (ET) \cite{Punturo:2010zza}) and neutrinos (LBNE \cite{Adams:2013qkq} and Hyper-KAMIOKANDE \cite{Abe:2011ts}) are implemented in the future, the sensitivity to both GW propagation speed and neutrino mass will be improved by extending the detection range of detectors up to $1\,{\rm{Mpc}}$. In Fig.~\ref{fig2b}, the detectable parameter ranges with a SN source at $1\,{\rm{Mpc}}$ are shown. In this case, detectable neutrino mass is about three times smaller and $m_{\nu} \geq 0.15\,{\rm{eV}}$ is detectable. However, since the cosmological constraint on the neutrino mass is $\sum m_{\nu} \leq 0.34\,{\rm{eV}}$, we can marginally neglect the neutrino mass uncertainty in constraining $\delta_g$, though the constraint can be corrected by a few tens of per cents. When the GW propagation is superluminal, $\delta_g$ is negative and does not calcel with $\delta_{\nu}$ in Eq.~(\ref{eq15}). However, we see above that the neutrino mass can be neglected in the realistic mass range. Then the superluminal propagation of a GW also gives the same results as the subluminal propagation.

Next let us see how the detectable range of $\delta_g$ changes depending on the intrinsic time delay of neutrinos and photons emission. To do so, we also consider SGRB as a potential source of the multimessenger observation of a GW, neutrinos, and photons. As for the prompt emission of SGRB, high energy photons are radiated in advance or behind the GW emission time. Compiling various models of long GRB, Baret et al. \cite{Baret:2011tk} found that the intrinsic time delay is in the range of $-150\,{\rm{sec}} < \tau_{\rm{int}} < 350\,{\rm{sec}}$, namely, $\Delta \tau_{\rm{int}}=500\,{\rm{sec}}$. For SGBR, this time window would be much smaller since the duration of the SGRB is typically less than $\sim 2\,{\rm{sec}}$. Then the expectation of the typical time delay from the point of view of dynamics would be $\sim 10\,{\rm{sec}}$. Therefore we use $\Delta \tau_{\rm{int}}=500\,{\rm{sec}}$ as a conservative bound and $\Delta \tau_{\rm{int}}=10\,{\rm{sec}}$ as a typical bound. Neutrinos are also emitted in SGRB and their emission time delay could be much shorter than that of photons \cite{Sekiguchi:2011zd}. However, the detection distance range and detection rate of neutrinos is similar to that in SN so that we omit the analysis for SGRB neutrinos. 

\begin{figure}[t]
\begin{center}
\includegraphics[width=8.5cm]{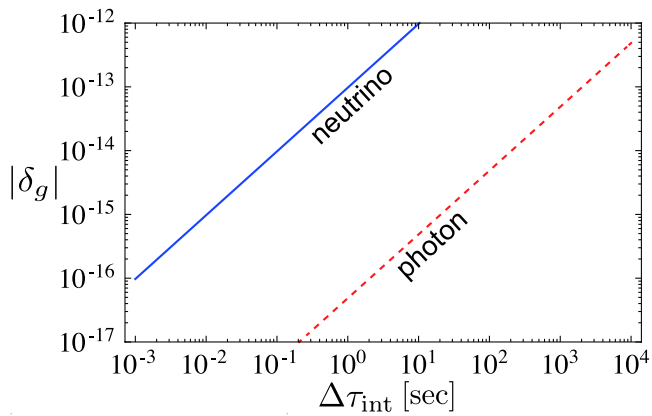}
\caption{Constraint on the propagation speed of a GW as a function of intrinsic time delay from multimessenger observations of a GW and, SN neutrinos (blue, solid) or SGRB photons (red, dashed). For SN, the neutrino energy is $10\,{\rm{MeV}}$ and the distance to the source is $100\,{\rm{kpc}}$. For SGRB, the distance to the source is $200\,{\rm{Mpc}}$.}
\label{fig3}
\end{center}
\end{figure}

In Fig.~\ref{fig3}, the constraint is shown as a function of $\Delta \tau_{\rm{int}}$. As discussed above, we choose $\Delta \tau_{\rm{int}}=10\,{\rm{msec}}$ as a typical time lag in the GW-SN observation. In the case of the null detection of a finite time lag in SN GW-neutrino observations, we have the constraint on $\delta_g$ for a SN event at $L=100\,{\rm{kpc}}$.
\begin{equation}
|\delta_g| < 9.7\times 10^{-16} \;. 
\label{eq14} 
\end{equation}
As for a SGRB, typical time lag is $\Delta \tau_{\rm{int}}=10\,{\rm{sec}}$ and conservative time lag is $\Delta \tau_{\rm{int}}=500\,{\rm{sec}}$. If the finite deviation of $\delta_g$ is not found in the GW-photon observations of a SGRB at $L=200\,{\rm{Mpc}}$, we would obtain the constraint on $\delta_g$; 
\begin{align}
|\delta_g| &< 2.4\times 10^{-14} \quad {\rm{for}}\;\; \Delta \tau_{\rm{int}}=500\,{\rm{sec}} \;, \label{eq16} \\
|\delta_g| &< 4.9\times 10^{-16} \quad {\rm{for}}\;\; \Delta \tau_{\rm{int}}=10\,{\rm{sec}} \;.
\label{eq17}
\end{align}
Since the constraint on $\delta_g$ is inversely proportional to $L$, if SGRB is associated with NS-BH binary of mass $1.4M_\odot$ and $10M_\odot$, respectively, the distance range is $\sim 3.4$ times larger \cite{Yonetoku:2014fua} so that the constraint would be improved by a factor $\sim 3$. In future, GW detectors (ET) and neutrino detectors (LBNE and Hyper-KAMIOKANDE) would be able to increase the constraint in Eqs.~(\ref{eq16}) and (\ref{eq17}) by about an order of magnitude.

The only direct measurement of the GW propagation speed proposed so far is a recent work by Finn and Romano \cite{Finn:2013toa}. Their method is based on the measurement of the R\o mer time delay, which modulates a periodic GW signal due to the Earth revolution. For a rapidly rotating nonaxisymmetric NS observed by a ground-based detector with ${\rm{SNR}}=10$, the constraint on the deviation of the GW propagation speed from speed of light is $|\delta_g| < 10^{-6}$. For galactic close white-dwarf binary systems observed by LISA-like detector with ${\rm{SNR}}=100$, the constraint is $|\delta_g| <10^{-3}$. The advantage of their method is that it does not necessarily require the electromagnetic counterpart, though indeed a priori knowledge about GW frequency and the sky location of a source helps improve the SNR. On the other hand, in our method, with a reasonable bound on the emission time delay, our constraints on the GW propagation speed are about 8-10 orders of magnitude tighter than the constraint from the R\o mer time delay. 

We also should comapre with the indirect constraint on $\delta_g$ obtained so far. From the measurement of ultra-high energy cosmic rays, assuming the cosmic rays come from galactic sources, the absence of energy loss due to gravitational Cherenkov radiation gives the constraint $0 \leq \delta_g < 2 \times 10^{-15}$ \cite{Moore:2001bv}. In the case of subluminal propagation, our method will give a stronger constraint by a factor of a few. The cosmic-ray constraint cannot be applied to the superluminal case, while our method gives the same constraint for both superluminal and subluminal cases. 

%%%%%%%%%%%%%%%%%%%%%%%
\subsection{Constraint on graviton mass}
The constraint on the propagation speed of a GW relies on the direct comparison of arrival times of a GW, neutrinos, and photons, and does not depend on the details of gravity theory. If one attributes the possible deviation of the GW propagation speed to graviton mass, the constraint on $\delta_g$ is converted to the constraint on the graviton mass. Substituting Eq.~(\ref{eq6}) into Eq.~(\ref{eq15}) and neglecting the neutrino mass uncertainty, we obtain the graviton mass limit in the absence of positive detection,
\begin{equation}
m_g < \sqrt{2}E_g \left( \frac{\Delta \tau_{\rm{int}}}{T_0}  \right)^{1/2} \;.
\end{equation}

With the choice of the intrinsic time delay, $10\,{\rm{msec}}$ for SN neutrinos and $500\,{\rm{sec}}$ ($10\,{\rm{sec}}$) for SGRB photons, we obtain the constraint $m_g < 1.8 \times 10^{-20}\,{\rm{eV}}$ and $m_g < 9.0 \times 10^{-20}\,{\rm{eV}}$ ($m_g < 1.3 \times 10^{-20}\,{\rm{eV}}$), respectively. Currently graviton mass has been constrained by several observations of the galaxy, the solar system, and binary pulsars (for summary, see \cite{Berti:2011jz} and references therein). However, the constraints from the galaxy and the solar system have been obtained from the observations in static gravitational fields and cannot be applied directly to GWs. While only the mass limit from binary pulsars comes from dynamical gravitational fields and is given by $m_g < 7.6\times 10^{-20}\,{\rm{eV}}$ \cite{Finn:2001qi}. Thus our bound on graviton mass with the propagation speed comparison is comparable or slightly stronger than the current bound from the binary pulsars. There have been proposals of the possible constraints on graviton mass from the future observation of a compact binary with a ground-based GW detector such as aLIGO. The observations of a NS binary and an intermediate mass black hole binary will give $m_g < 2.7\times 10^{-22}\,{\rm{eV}}$ \cite{Will:1997bb} and $m_g < 1.6\times 10^{-23}\,{\rm{eV}}$ \cite{Arun:2009pq,Keppel:2010qu}, respectively. These bounds are much tighter than our bound because the theoretical GW templates including the phase shift due to graviton mass are used for the detection. However, it is well known that a simple introduction of the mass term in the metric theory arises difficulties shown by Boulware and Deser \cite{Boulware:1973my}. To argue the massive gravity, one possible theory is the ghost free massive bi-gravity theory \cite{deRham:2010ik,deRham:2010kj,Hassan:2011hr}. In this theory, the graviton speed depends not only mass but also model parameters and graviton oscillation should also be taken into account as shown for example in \cite{DeFelice:2013nba}.

\subsection{Event rate}
We comment on the event rate of coincidence detections of a GW and, neutrinos from SN or photons from SGRB. In SN GW-neutrino observations, the detectable distance is limited by neutrino detectors to be within $100\,{\rm{kpc}}$, which gives the coincident event rate of roughly a few events per a century \cite{Ando:2005ka}. On the other hand, in SGRB GW-$\gamma$-ray detectors, the GW detectors limits the detectable distance to be within $200\,{\rm{Mpc}}$and $700\,{\rm{Mpc}}$ for NS-NS case and NS-BH case, respectively. Recently Yonetoku {\it{et al.}} \cite{Yonetoku:2014fua} estimated the minimum event rate of SGRB per co-moving volume using BATSE data as $R_{\rm on-axis}^{\rm min} = 6.3_{-3.9}^{+3.1} \times 10^{-10}~{\rm events~Mpc^{-3}yr^{-1}}$. This  corresponds to the minimum GRB GW-photons coincident event rate of $\sim 0.02\,{\rm{yr}}^{-1}$ and $\sim 0.9\,{\rm{yr}}^{-1}$ for NS-NS case and NS-BH case, respectively. It is also argued in \cite{Yonetoku:2014fua} that the true event rate would be $\sim$ four times larger so that they would be $\sim 0.08\,{\rm{yr}}^{-1}$ and $\sim 3.6\,{\rm{yr}}^{-1}$ for NS-NS case and NS-BH case, respectively. The rate for the NS-NS case is small but that for the NS-BH case is large enough and the NS-NS case is comparable to that of the SN GW-neutrinos event rate. Thus we can expect that we will have at least one SGRB coincident event after from a few to several years observation. Note here that if photons from off-axis SGRB are detected, the event rate will increase by a factor $\sim 200$ \cite{Yonetoku:2014fua}, which is similar to that estimated by \cite{Abadie:2010cf}, so that the coincidence event rate for the NS-NS case is much higher than that of the SN GW-neutrinos event rate. Finally, we notice that in any case it is possible to have a fortune SN coincident event before long observations.

\section{Distinguishing signal and noise with multiple cosmological SGRB}
\label{sec4}

In the discussion in Sec.~\ref{sec3}, the distance to a SN or SGRB source is limited by the detection ranges of GW and neutrino detectors. The number of a coincidence event is limited to be one or a few in realistic observation time. However, the future ground-based GW detector, ET, extends the detection range by more than ten times and enables us to observe the merger events of NS-NS binaries at cosmological distance up to $z\sim2$, while for NS-BH binaries up to $z\sim 4$ \cite{Sathyaprakash:2009xt}. As discussed in \cite{Nishizawa:2010xx} and Refs. therein, ET will detect a million of compact binaries. To utilize them for our purpose here, we also need the detections of SGRB as an electromagnetic counterpart. Even in a pessimistic case, from the consideration of the beaming angle of SGRB, more than 100 GW-SGRB coincidence events would be observed in a realistic observation time, e.g. 3\,{\rm{yr}}. In this section, we discuss how the time delay is affected by the cosmic expansion and how the constraint on $\delta_g$ changes. Also we discuss a possible method to distinguish the time delay due to $\delta_g$ from the intrinsic time delay with multiple SGRB at cosmological distance.

Let us assume a flat $\Lambda$CDM universe for simplicity. The comoving distance to a source at redshift $z$ is
\begin{align}
\chi (z) &= \int_{0}^{z} \frac{\upsilon}{H(z) a_0}dz \nonumber \\
&= (1-\delta) \chi_0 (z) \;,
\end{align}
Here we assume $v$ is independent of the redshift and defined $\delta \equiv (c-\upsilon)/c$. $a_0$ is a scale factor at present and $\chi_0 (z)$ is the comoving distance when $\delta_g=0$. $H(z)$ is the Hubble parameter given by
\begin{equation}
H(z) = H_0 \sqrt{\Omega_{\rm{m}}(1+z)^3+\Omega_{\Lambda}} \;.
\end{equation}
where $\Omega_{\rm{m}}$ and $\Omega_{\Lambda}=1-\Omega_{\rm{m}}$ are the energy densities of matter and a cosmological constant, and $H_0$ is the Hubble constant at present. We take the values $H_0=100\, h_0 \,{\rm{km}}\,{\rm{Mpc}}^{-1}\,{\rm{s}}^{-1}$ with $h_0=0.68$ and $h_0^2 \Omega_{\rm{m}}=0.14$ \cite{Ade:2013zuv}. 

To derive the time delay induced by $\delta_g$ for a source at cosmological distance,
we compare the particles emitted from the same source but with different velocities, $c$ and $\upsilon_g$. Both particles are emitted at the same redshift $z$ but reaches us at $z=0$ and $z=-\Delta z$, respectively. The comoving distance from the source to the Earth is the same for both particles. Neglecting the second-order correction in $\Delta z$, $\chi(z+\Delta z)=\chi_0(z)$ gives  
\begin{equation}
\Delta z = H_0 \delta_g \int_{0}^{z} \frac{dz^{\prime}}{H(z^{\prime})} \;.
\end{equation} 
Thus the time delay induced by $\delta_g$ is
\begin{equation}
\Delta T = \frac{\Delta z}{H_0}= \delta_g \int_{0}^{z} \frac{dz^{\prime}}{H(z^{\prime})} \;. \label{eq4}
\end{equation} 

Also the intrinsic time delay is redshifted. Denoting the intrinsic time delay at the source as $\tau_{\rm{int}}^{(\rm{e})}$, the time delay we observe at the Earth is
\begin{equation}
\tau_{\rm{int}} =(1+z)\, \tau_{\rm{int}}^{(\rm{e})} \;.
\end{equation} 

\begin{figure}[t]
\begin{center}
\includegraphics[width=8.5cm]{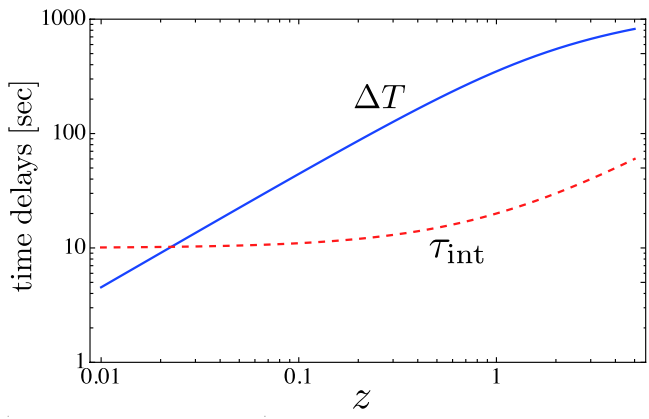}
\caption{Arrival time lag between a GW and a photon (blue, solid) and the intrinsic time delay (red, dashed) of a cosmological SGRB as a function of redshift. For illustration, the parameters are chosen as $\delta_g=10^{-15}$ and $\tau_{\rm{int}}^{(\rm{e})}=10\,{\rm{sec}}$.}
\label{fig4}
\end{center}
\end{figure}

In Fig.~\ref{fig4}, the time delays due to $\delta_g$ and emission mechanism are illustrated for the case of $\delta_{g}=10^{-15}$ and $\tau_{\rm{int}}^{(\rm{e})}=10\,{\rm{sec}}$. As expected from Eqs.~(\ref{eq3}) and (\ref{eq4}), the time delay due to finite $\delta_g$ increases at low $z$, proportional to the distance to the source. However, at high $z$, the cosmic expansion changes from acceleration to deceleration, which modifies the dependence of the time delay on the distance (redshift). So the growth of the time delay at high $z$ slows down and starts to decrease, being proportional to $z^{-1/2}$. On the other hand, the intrinsic time delay is redshifted but almost constant at low $z$. At high $z$, it increases proportional to $z$. Thus the SNR is proportional to $z$ at low $z$ and $z^{-3/2}$ at high $z$. 

\begin{figure}[t]
\begin{center}
\includegraphics[width=8.5cm]{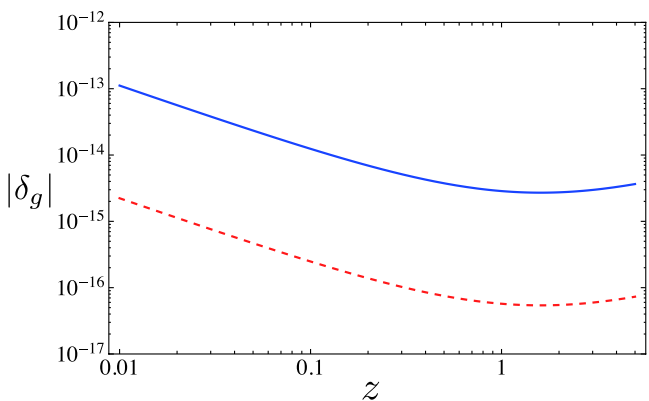}
\caption{Constraint on $\delta_g$ from a single SGRB coincidence event as a function of redshift. The intrinsic time delays are $\Delta \tau_{\rm{int}}=500\,{\rm{sec}}$ (blue, solid) and $\Delta \tau_{\rm{int}}=10\,{\rm{sec}}$ (red, dashed).}
\label{fig5}
\end{center}
\end{figure}

In the absence of a positive signal, we can constrain $\delta_g$, using Eq.(\ref{eq15a}). In Fig.~\ref{fig5}, we show the possible constraint on $\delta_g$ with a  SGRB event at cosmological distance, assuming typical and conservative cases, $\Delta \tau_{\rm{int}}^{(\rm{e})}=10\,{\rm{sec}}$ and $\Delta \tau_{\rm{int}}^{(\rm{e})}=500\,{\rm{sec}}$. It is interesting that the best sensitivity is obtained from the source at $z=1-3$. With a source at $z=1$, we obtain the constraint
\begin{align}
|\delta_g| &< 2.9\times 10^{-15} \quad {\rm{for}}\;\; \Delta \tau_{\rm{int}}=500\,{\rm{sec}} \;, \\
|\delta_g| &< 5.7\times 10^{-17} \quad {\rm{for}}\;\; \Delta \tau_{\rm{int}}=10\,{\rm{sec}} \;.
\end{align}

If one has multiple SGRB events observed coincidentally by GW and $\gamma$-ray detectors, one can distinguish the true signal due to the deviation of GW propagation speed from the speed of light and the intrinsic time delay of the emission at a source by looking at the redshift dependence. To do so, we propose a new statistics that can be used in the data analysis. The observed quantity is $\tau_{\rm{obs}}$ in Eq.~(\ref{eq12}), from which we can construct the following statistic:
\begin{align}
\Delta \tau_{\rm{obs}} (z_i,z_j) &\equiv \frac{\tau_{\rm{obs}} (z_i)}{1+z_i} - \frac{\tau_{\rm{obs}} (z_j)}{1+z_j} \nonumber \\
&= \frac{\Delta T(z_i)}{1+z_i}-\frac{\Delta T(z_j)}{1+z_j} + \tau_{\rm{int}}^{(\rm{e})}(z_i)-\tau_{\rm{int}}^{(\rm{e})}(z_j) \;. \label{eq7}
\end{align}
Since $\tau_{\rm{int}}^{(\rm{e})}$ is expected to be distributed about its average, depending on a specific model of SGRB emission, the difference of the third and fourth terms would be a stochastic noise and approach zero on average over many data samples. Then summing the signals over $(i,j)$ pairs of redshifts with $z_j < z_i$, we expect Eq.~(\ref{eq7}) to approach
\begin{align}
\sum_{z_j<z_i} \Delta \tau_{\rm{obs}} (z_i,z_j) &= \sum_{z_j<z_i} \left\{ \frac{\Delta T(z_i)}{1+z_i}-\frac{\Delta T(z_j)}{1+z_j} \right\} \nonumber \\
&\sim (N-1) \frac{\Delta T(z_{\rm{max}})}{1+z_{\rm{max}}} \;.
\end{align}
where $N$ is the number of SGRB sources. The second line is obtained from the rough estime that $\Delta T$ at high $z$ dominates as shown in Fig.~\ref{fig4}. Further detailed evaluation of this statistic needs numerical investigation, including the noise power spectrum of a specific detector, the number of coincidentally observed events by GW and $\gamma$-ray detectors, and the redshift distribution of the NS-NS binary merger rate, etc. So we leave it as the future work.

%%%%%%%%%%%%%%%%%%%%%%%%%%%%%%%%%%%
\section{Conclusions and discussions}
We have proposed the method measuring the propagation speed of a GW by directly comparing the arrival time lags between a GW and, neutrinos from SN or photons from SGRB. We have found that the multimessenger observations of GWs from nearby SN and SGRB enable us to constrain the deviation of the GW propagation speed from the speed of light to be $|\delta_g| < 9.7\times 10^{-16}$ for SN and $|\delta_g| < 4.9\times 10^{-16}$ for SGRB, respectively, improving the sensitivity of the previous method using R\o mer time delay by 8-10 orders of magnitude. We also have shown that with ET the above constraint will be improved by an order of magnitude or more by observing multiple sources at cosmological distance. Importantly, one can distinguish the true signal due to the deviation of GW propagation speed from the speed of light and the intrinsic time delay of the emission at a source by looking at the redshift dependence. This enables us to use our method as a more robust test of gravity.

Finally, we comment on the possible errors in a realistic observation. In this paper, we have simplified our analysis by assuming that the distance to the source is known accurately. However, in realistic experiments, it is not easy to obtain the distance information and its uncertainty could be a systematic error. If the redshift is identified from an electromagnetic spectrum, the distance error would be $1\%$ or less \cite{Dietz:2010eh} and negligibly small. If not identified, the distance error of the GW observation is $20-30\%$ or less for aLIGO within their detection range of $\sim 250\,{\rm{Mpc}}$ \cite{Nissanke:2009kt} and $10\%$ or less for ET within $z=3$ \cite{Camera:2013xfa}. Then even in the worse case, our constraints would be degraded by up to $30\%$. We also have to consider the statistical error of the arrival time that arises from small samples of detected neutrinos or photons. Assuming the distribution is Poissonian and requiring the detection of at least one particle at 2-$\sigma$ level, we need $N \geq 5.8$ samples to claim the positive detection. Although the statistical error of the arrival time does not change our results qualitatively, further studies are necessary including the realistic flux of the particles and the detection efficiency of the detectors.

\begin{acknowledgments}%%%%%%%%%%%%%%%%%%%%%%%%%%%%%%%
We would like to thank S.~Desai, I. Sawicki, M.~Sissa, Y.~Suwa, and T.~Yokozawa for valuable comments. A.N. and T. N. are supported by Grant-in-Aid for Scientific Research on Innovative Areas, No.24103006. Also A. N. is supported by JSPS Postdoctoral Fellowships for Research Abroad and T. N. is supported by Grants-in-Aid for Scientific Research No.23540305.
\end{acknowledgments}

\bibliography{/Volumes/USB-MEMORY/my-research/bibliography}

\end{document}